\documentclass{article}

\usepackage{arxiv}
\usepackage[utf8]{inputenc} 
\usepackage[T1]{fontenc}    
\usepackage{hyperref}       
\usepackage{url}            
\usepackage{booktabs}       
\usepackage{amsfonts}       
\usepackage{nicefrac}       
\usepackage{microtype}      
\usepackage{lipsum}		
\usepackage{natbib}
\usepackage{doi}
\usepackage{amsmath,amssymb,amsthm,amsfonts}
\usepackage{algorithmic,algorithm, array}
\usepackage{textcomp,stfloats,verbatim}
\usepackage{graphicx,xcolor,tikz}
\usepackage{authblk, subcaption}
\usetikzlibrary{patterns}

\newcommand{\winlength}{\theta} 
\newcommand{\standardwinlength}{L} 
\newcommand{\support}{N} 

\newcommand{\complex}{j}

\newcommand{\idx}{t}
\newcommand{\col}{i} 
\newcommand{\f}{\xi} 

\newcommand{\loss}{\mathcal{L}}
\newcommand{\ent}{\mathcal{H}}

\newcommand{\spec}{\mathcal{S}}

\definecolor{darkred}{rgb}{.82,0,0}
\definecolor{BleuBleu}{RGB}{0 50 150}
\definecolor{darkgreen}{rgb}{0,.4,0}
\definecolor{lmcolor}{rgb}{0,.6,0}
\definecolor{ymcolor}{rgb}{1,.4,0}
\definecolor{abcolor}{rgb}{0,.6,.9}
\definecolor{mlcolor}{rgb}{0.5,.2,.5}

\title{Differentiable adaptive short-time Fourier transform with respect to the window length}

\date{} 					

\author{ Maxime Leiber{$^{* \dagger}$}, Yosra Marnissi{$^\dagger$}, Axel Barrau{$^\ddagger$}, Mohammed El Badaoui{$^\dagger \mathsection$} \vspace{.5cm} \\
 	{$^*$} INRIA, DI/ENS, PSL Research University \\
 	{$^\dagger$} Safran Tech, Digital Sciences \& Technologies \\
 	{$^\ddagger$} Offroad \\
        {$^\mathsection$} Univ Lyon, UJM-St-Etienne, LASPI \\
}

\renewcommand{\headeright}{}
\renewcommand{\undertitle}{Accepted for IEEE ICASSP 2023}
\renewcommand{\shorttitle}{Differentiable short-time Fourier transform}

\hypersetup{
pdftitle={Differentiable short-time Fourier transform},
pdfsubject={Differentiable short-time Fourier transform},
pdfauthor={Maxime Leiber},
pdfkeywords={Time-frequency, differentiable STFT, adaptive STFT, spectrogram, gradient descent},
}

\begin{document}
\maketitle

\begin{abstract}
This paper presents a gradient-based method for on-the-fly optimization for both per-frame and per-frequency window length of the short-time Fourier transform (STFT), related to previous work in which we developed a differentiable version of STFT by making the window length a continuous parameter. 
The resulting differentiable adaptive STFT possesses commendable properties, such as the ability to adapt in the same time-frequency representation to both transient and stationary components, while being easily optimized by gradient descent. We validate the performance of our method in vibration analysis.
\end{abstract}


\section{Introduction}
\label{sec:1_}

Fourier theory is a crucial aspect of signal processing, widely used in science and engineering. The short-time Fourier transform (STFT), also known as the windowed Fourier transform, plays a vital role in analyzing non-stationary signals with time-varying spectral content. Spectrograms, derived from the STFT magnitude, are commonly used for visualizing and processing non-stationary signals. The STFT window length is a critical parameter that determines the trade-off between temporal and frequency resolution, and several post-processing techniques have been developed to improve spectrogram readability, including synchrosqueezing \cite{Thakur} and reassignment \cite{Auger}. 
Some researchers have proposed finding the optimal window length based on a given criterion \cite{Meignen, Intelligent}, while others have recently proposed a differentiable version of STFT with respect to the window length\cite{dstft, gretsi, Zhao}, allowing for the optimization of the criterion using a gradient descent algorithm instead of grid search.

Actually, the best window length depends on the signal itself and more particularly on its frequency content. It must therefore adapt to the time-varying spectral structure of the signal. Enhanced versions of STFT are then proposed to set the window length according to the local characteristics of the input signal. These methods are known as adaptive STFTs \cite{Czerw, Kwok,Zhong,Pei,Zhu}. They use a different window length per frame and per frequency. The optimal values for the window lengths in the 2D plane are chosen to favor a given criterion such as sparsity \cite{Pei} or local stationarity \cite{explorations}.

Our main contribution is to provide a differentiable version of adaptive STFT where window lengths can be easily optimized per-frame and per-frequency using gradient descent. This paper is organized as follows. In Section \ref{sec:2_} we first give some definitions and notations of STFT, differentiable STFT and adaptive STFT. In Section \ref{sec:3_} we introduce our modified differentiable adaptive STFT w.r.t. the window length and propose an optimization criterion. In Section \ref{sec:4_} we demonstrate the effectiveness of our approach with a simulated illustration and an experiment applied to vibration analysis in aeronautics and we end our discussion in Section \ref{sec:5_} with some final remarks. 

\section{Definitions and related works}
\label{sec:2_}
\subsection{Short-Time Fourier Transform}
All over this paper we will refer to STFT as the operation taking a one-dimensional signal $s[t]$ as input and returning a one-dimensional matrix $\spec[\col,\f]$. Each column $\spec[\col,:]$ of the STFT is the Discrete Fourier Transform (DFT) of a slice of length $\standardwinlength$ of the signal $s$ starting from an index $\idx_\col$ to an index $\idx_\col + \standardwinlength-1$, multiplied by a \emph{tapering function} $g_\standardwinlength$ of length $\standardwinlength$. STFT can be mathematically written as follows:
\begin{equation}
\label{eq::eq1}
\begin{aligned}
\spec[\col,f] 
 & = \sum_{k=0}^{\standardwinlength-1} g_\standardwinlength[k]s[\idx_\col+k] e^{-2\complex\pi kf/\standardwinlength}.
 \end{aligned}
\end{equation}
Starting indices $b_i$ of time intervals on which spectra are computed are usually equally spaced, so we only have to set the first index $\idx_0$ and spacing $\Delta \idx$ between $\idx_i$ and $\idx_{i+1}$. Finally, several choices of tapering function $g_\standardwinlength$ can be encountered, such as the Gaussian and the Hann windows.

\subsection{Differentiable Short-Time Fourier Transform}
Tuning the window length is crucial to ensure a good time-frequency resolution whereas this hyperparameter is usually fixed empirically by trial-and-error. A differentiable version of STFT (DSTFT) has been recently proposed in \cite{dstft}. DSTFT modifies the STFT operator to make the window length a continuous parameter w.r.t. which spectrogram values can be easily differentiated. This differentiable STFT can be integrated easily into any existing algorithm (e.g. neural networks) involving spectrograms and the window length can be optimized for a given cost function (e.g. neural network loss) using gradient descent. The idea behind differentiable STFT is to break down the window length $\standardwinlength$ into an \emph{integer numerical window support} and a \emph{continuous time resolution}.

\subsection{Adaptive Short-Time Fourier Transform}
Classical STFT uses a single window for each bin. 
However, this is very limited for signals with time-varying spectral content. In fact, it seems more appropriate for such signals to distinguish regions with transient content where a small window length is required from regions where stationary activity is prominent where longer windows are preferred. One known representation is the S-transform that uses a decreasing window length along frequencies bins \cite{Stockwell}.
More general adaptive STFT (ASTFT) have been proposed in the literature. They use a different window for each frame $\col$ and frequency $f$ among a set of windows of varying lengths. The adaptive STFT can be defined as:  
\begin{equation}
\label{eq::eq3}
\begin{aligned}
\spec[\col,f] = \sum_{k=0}^{\standardwinlength-1} g_{\standardwinlength_{i,f}}[k]s[\idx_\col+k] e^{-2\complex\pi kf/\standardwinlength}.
 \end{aligned}
\end{equation}
To our knowledge, all proposed ASTFT use the Gaussian window and aim to adapt the variance parameter to each bin. This last parameter is commonly selected by minimizing an adaptation criterion $\loss$ related to uncertainty, such as local variance \cite{explorations}, energy concentration or sparsity \cite{Pei}. The minimization is often performed by grid search \cite{Intelligent}. Few works have proposed iterative numerical algorithms where a prior estimation of the instantaneous frequency is usually required. The methods are not detailed here since our goal is to propose an ASTFT that can be optimized by gradient-based algorithms.

\section{Differentiable adpative short-time Fourier transform}
\label{sec:3_}
\subsection{Mathematical formulation}

Defining an adaptive version of differentiable STFT amounts to writing a formula similar to that of \cite{dstft} where time resolution $\winlength$ varies with frame $\col$ and frequency $f$:
\begin{equation}
\label{eq::eq6}
\spec[i,f] = \sum_{k=0}^{\support-1} g_{\support,\winlength_{if}}[k] s[b_i+k] e^{-2\complex\pi kf/\support}.
\end{equation}
In \eqref{eq::eq6}, $N$ refers to the numerical window support that can been seen as an upper bound of the continuous time resolution parameter $\winlength_{if}$. In the following, we will denote by $\Theta$ the 2D matrix formed of continuous time resolutions parameters  $\winlength_{if}$. 

Let us now compute the differential of our proposed STFT w.r.t. to $\winlength_{if}$. $\spec[\col,f]$ being complex, we apply the term-by-term differentiation, by considering complex numbers as vectors with two real components, in particular $\exp{(jx)} = \left[\cos(x), \sin(x) \right]$. We obtain the Jacobian of size (2, 1):
\begin{equation}
\label{eq::eq7}
\frac{\partial \spec[\col,f]}{\partial \winlength_{if}} = \sum_{k=0}^{\support-1} \exp{ \left(-2\complex \pi \frac{kf}{\support}\right)} \frac{\partial g(k)}{\partial \winlength_{if}} s[\idx_\col+k]
\end{equation}
where we recognize the STFT of $s$ with tapering function $g_{if}'= \frac{\partial g_{\support, \winlength_{if}}(k)}{\partial \winlength_{if}}$ instead of $g_{\support, \winlength_{if}}$:
\begin{equation}
\frac{\partial \spec[s]}{\partial \winlength_{if}} = \spec_{g_{if}'}[s]
\end{equation}
The latter result allows deriving compact gradient backpropagation formulas for gradient descent based optimization algorithms. In particular, given the adaptation criterion $\loss$, we directly obtain from \eqref{eq::eq6}:
\begin{equation}
\label{eq::eq9}
\frac{\partial \loss}{\partial \winlength_{if}} = \frac{\partial \loss}{\partial \spec[\col,f]} \frac{\partial \spec[\col,f]}{\partial \winlength_{if}}
= \frac{\partial \loss}{\partial \spec[\col,f]} \spec_{g_{if}'}[s]
\end{equation}
where $\frac{\partial \loss}{\partial \spec(\col,f)}$ is the vector of derivatives w.r.t. real and imaginary parts of size (1, 2). Note that unlike previous work \cite{dstft}, there is no expression for this formula as a Froebenius scalar product.


 
\subsection{Optimization criterion}
We need a criterion to optimize $\winlength_{if}$. It is known that a good time-frequency representation minimizes the uncertainty on time and frequency \cite{explorations} and so, promotes parsimony. 
Several functions were used to promote sparse spectograms such as  statistical moments e.g. variance \cite{explorations}, kurtosis \cite{Zhao} or relative standard deviation \cite{Intelligent}, (quasi-)norm ratios $\mathbb{L}_p$-over-$\mathbb{L}_q$ \cite{Pei} and the Renyi entropy \cite{Meignen}.
After a non-exhaustive study, we have chosen the Shannon's entropy due to its popularity although all candidates had roughly equivalent results. The entropy loss function can be then written as:
\begin{equation}
\label{eq::eq10}
\ent(\vert \spec\vert |\Theta) = - \sum_{\col,f} p_{\col f} \log (p_{\col f})
\end{equation}
where $p_{if} = \frac{|\spec[\col, f]|}{\sum_{k, l} |\spec[k, l]|}$. 

Since real signals are corrupted with high levels of non-stationary noise in many domains such as aeronautics, it is necessary to impose some constraints on the windows length to ensure that the optimized windows are not disturbed by the presence of transient noise. Inspired by the work on image denoising, we propose to add the following regularization:
\begin{equation}
    \mathcal{R}(\Theta) = \sum_{i,f} \sqrt{\sum_{(j,k) \in V_{i,f}} \omega_{(j,k);(i,f)}(\winlength_{i,f}-\winlength_{j,k})^2} 
\end{equation}
where $V_{i,f}$ is a neighboring index set for the bin $(i,f)$ and $\omega_{j,k}$ is a weight given to the neighbor $\winlength_{j,k}$ to measure the similarity between the frequency content of the regions $(i,f)$ and $(j,k)$. These weights can be equal or set from a previously computed spectrogram with a constant window length. The above regularization is related to the non-local total variation penalization used in image processing \cite{nltv}. A simple variant, named the total variation, consists in considering the neighborhood as $V_{i,f}=\{(i+1, f), (i, f+1)\}$ with equal weights. Our regularization has the benefit of constraining the window lengths of (close) regions with similar spectral content to be close, which improves robustness to noise. The advantages of adding such a regularization will be further demonstrated in the illustrative example. 
We finally obtain the following optimization criterion:
\begin{equation}
\label{eq::eq12}
\loss(\Theta) = \ent(\vert \spec\vert|\Theta) + \lambda \mathcal{R}(\Theta)
\end{equation}
where $\lambda$ is an hyperparameter controlling trade-off between the two terms.

\subsection{Discussion}
The advantages of our proposed differentiable adaptive STFT (DASTFT) is threefold. First, it adapts the window length to the frequency content of the signal. Second, it is easily optimizable by gradient descent. Third, unlike previous works, it is not limited to the Gaussian window but can be computed with any differentiable tapering function \cite{dstft}.

\section{Applications to vibration analysis}
\label{sec:4_}
We now demonstrate our method through two examples motivated by vibration analysis. For all spectrograms, we use the Hann window. 


\subsection{Illustrative example}

\begin{figure}[h]
\centering
\begin{subfigure}{.33\textwidth}
  \centering 
  \includegraphics[width=5cm]{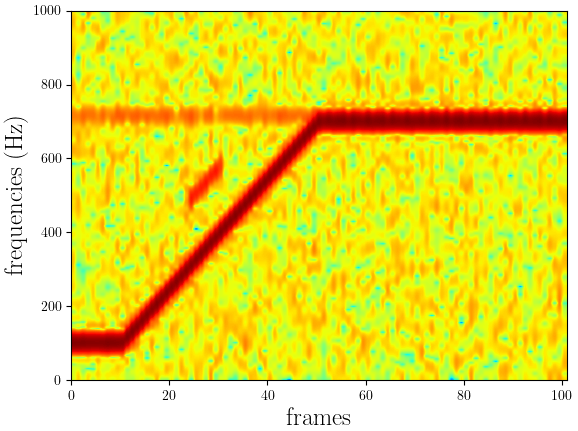}
  \caption{Window of length 100.}
\end{subfigure}%
\hspace{-.5cm} \begin{subfigure}{.33\textwidth}
  \centering 
  \includegraphics[width=5cm]{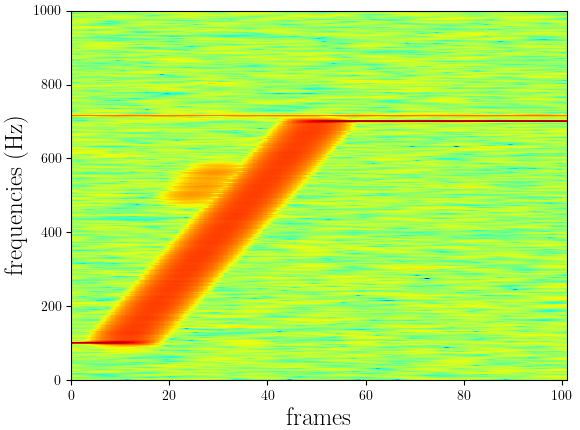}
  \caption{Window of length 1000.}
\end{subfigure}
\hspace{-.5cm} \begin{subfigure}{.33\textwidth}
  \centering 
  \includegraphics[width=5cm]{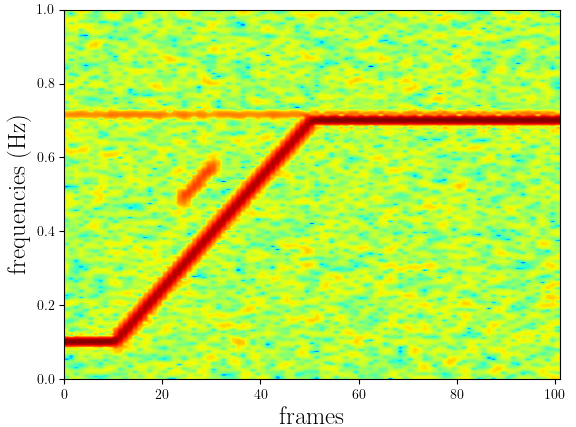}
  \caption{DSTFT with a single window length.}
\end{subfigure}
\caption{Spectrograms with different window length.}
\label{fig:fig1}
\end{figure}

We show the interest of DASTFT on a simulated signal representing a typical example of a vibration signal that can be measured in an aircraft engine\footnote{We have shared the notebook to reproduce the experiment at \url{https://github.com/maxime-leiber/dstft}}. The signal contains 3 harmonics. The main component is non-stationary, the second one is a close and constant interfering frequency which can represent a modulation and the third one is a transient event like an explosive fan flutter. We also added some noise. A small window length gives fine temporal resolution to localize transient events in time but a coarse frequency resolution to distinguish nearby frequencies and vice versa as displayed in Fig. \ref{fig:fig1} a) and b).  
DSTFT presented in \cite{dstft} gives an interesting compromise between time and frequency resolution. But since we use only one window for the whole spectrogram, we are not able to localize precisely in time transient event and in frequency close frequencies as shown in Fig.\ref{fig:fig1} c). 

\begin{figure}[t]
\centering
\begin{subfigure}{.25\textwidth}
  \centering 
  \includegraphics[width=4cm]{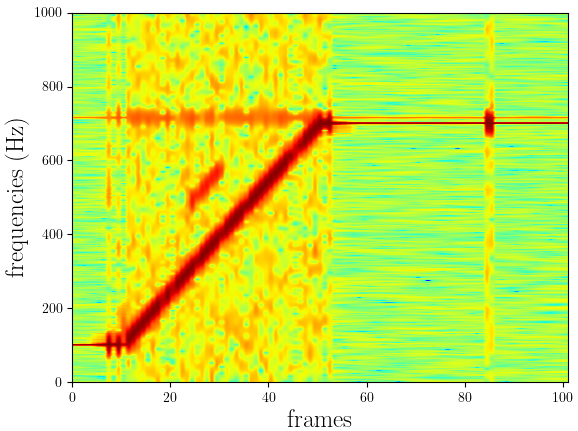}
  \caption{Spectrogram.}
\end{subfigure}%
\hspace{-.5cm} \begin{subfigure}{.25\textwidth}
  \centering 
  \includegraphics[width=3.7cm]{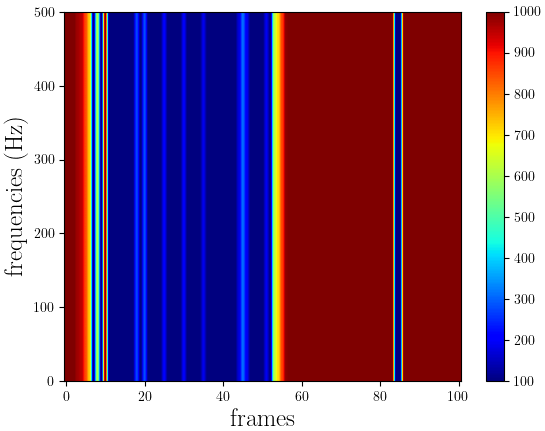}
  \caption{Windows length.}
\end{subfigure}
\centering
\begin{subfigure}{.25\textwidth}
  \centering 
  \includegraphics[width=4cm]{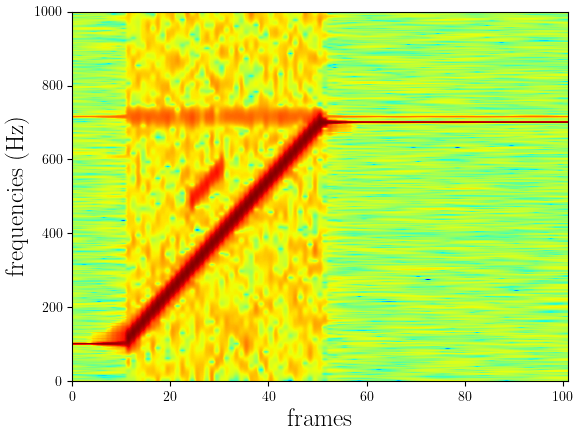}
  \caption{Spectrogram.}
\end{subfigure}%
\hspace{-.5cm} \begin{subfigure}{.25\textwidth}
  \centering 
  \includegraphics[width=3.7cm]{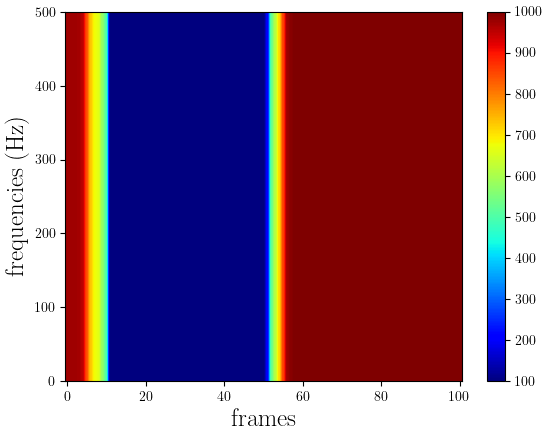}
  \caption{Windows length.}
\end{subfigure}
\caption{DASTFT with a time-varying window length without (top) and with (bottom) regularization.}
\label{fig:fig3}
\end{figure}

\begin{figure}[t]
\centering
\begin{subfigure}{.25\textwidth}
  \centering 
  \includegraphics[width=4cm]{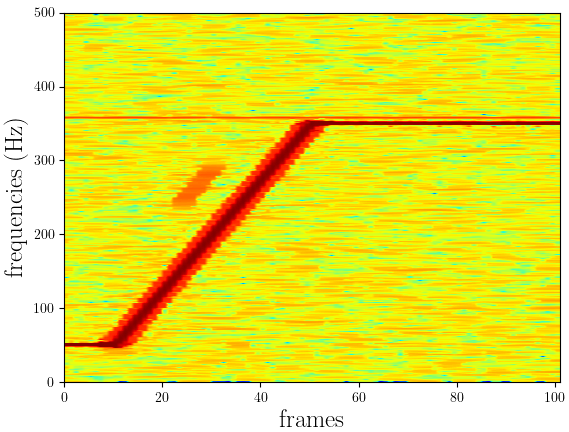}
  \caption{Spectrogram.}
\end{subfigure}%
\hspace{-.5cm} \begin{subfigure}{.25\textwidth}
  \centering 
  \includegraphics[width=3.7cm]{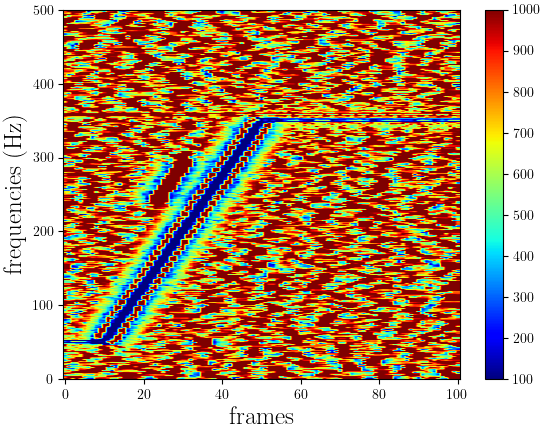}
  \caption{Windows length.}
\end{subfigure}
\begin{subfigure}{.25\textwidth}
  \centering 
  \includegraphics[width=4cm]{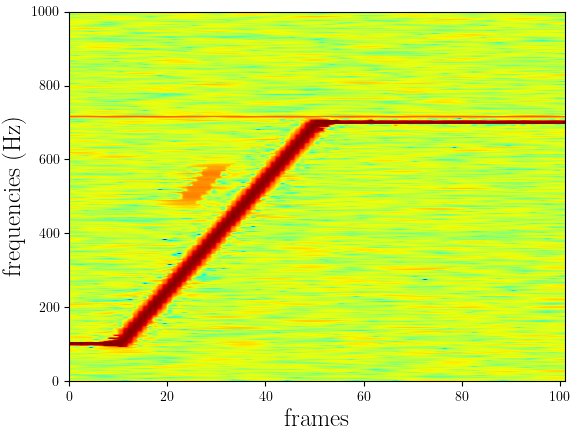}
  \caption{Spectrogram.}
\end{subfigure}%
\hspace{-.5cm} \begin{subfigure}{.25\textwidth}
  \centering 
  \includegraphics[width=3.7cm]{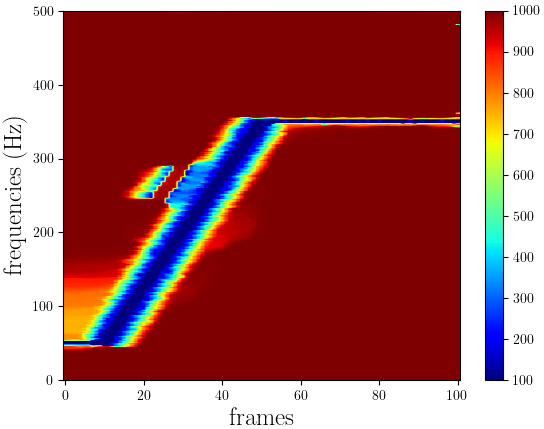}
  \caption{Windows length.}
\end{subfigure}
\caption{DASTFT with a time-and-frequency-varying window length without (top) and with (bottom) regularization.}
\label{fig:fig4}
\end{figure}

We now compute our DASTFT, which also allows gradient descent. In particular, we compute it with and without regularization in the adaptation loss. Results (spectrograms and window length distribution) are shown for only-time-varying window length and with time-and-frequency-varying window length in Fig.\ref{fig:fig3} and Fig.\ref{fig:fig4} respectively. We see that we can localize precisely all the components of our signal both in time and in frequency especially with time-and-frequency-varying window length. In fact, the latter adapts to each component as expected e.g. small windows for transient events and long windows for stationary components. However, we notice that the result is sensitive to noise when ony the entropy loss is considered. In Fig. \ref{fig:fig3}, artifacts appear due to an abrupt change of the window length in the stationary regions while in Fig. \ref{fig:fig4}, we see a high variability of the window length in regions containing only noise. These problems are solved by adding the proposed regularization term.

\subsection{Multi-harmonic vibration signal}

\begin{figure}[H]
\centering
\begin{subfigure}{.25\textwidth}
  \centering 
  \includegraphics[width=4cm]{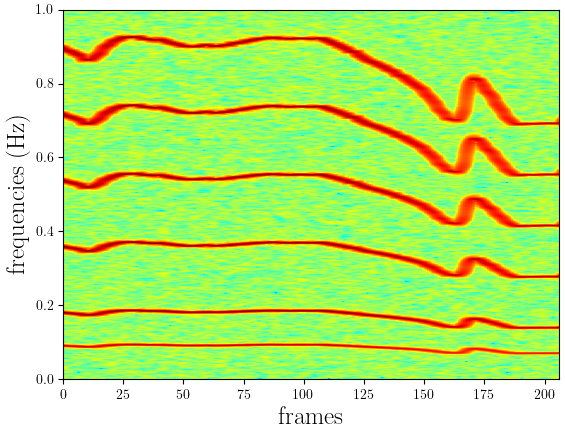}
  \caption{DSTFT.}
\end{subfigure}%
\begin{subfigure}{.25\textwidth}
  \centering 
  \includegraphics[width=4cm]{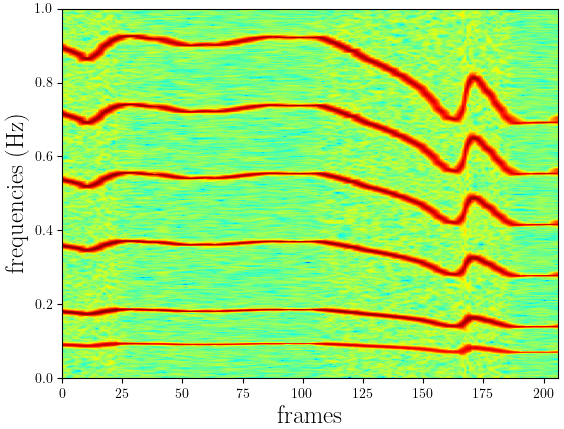}
  \caption{Spectrogram.}
\end{subfigure}%
\hspace{-.5cm} \begin{subfigure}{.25\textwidth}
  \centering 
  \includegraphics[width=3.8cm]{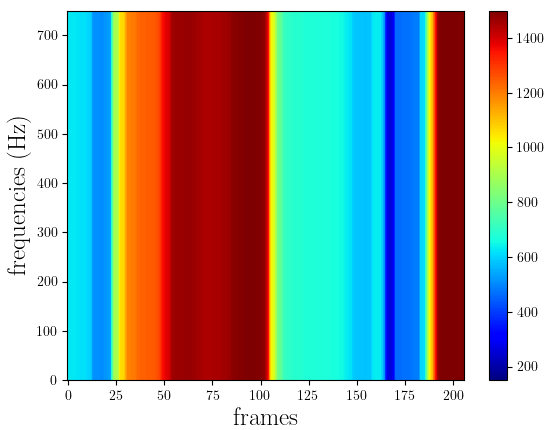}
  \caption{Windows length.}
\end{subfigure}
\caption{DASTFT with a time-varying window length.}
\label{fig::4.2.1}
\end{figure}

\begin{figure}[H]
\centering
\begin{subfigure}{.3\textwidth}
  \centering 
  \includegraphics[width=5cm]{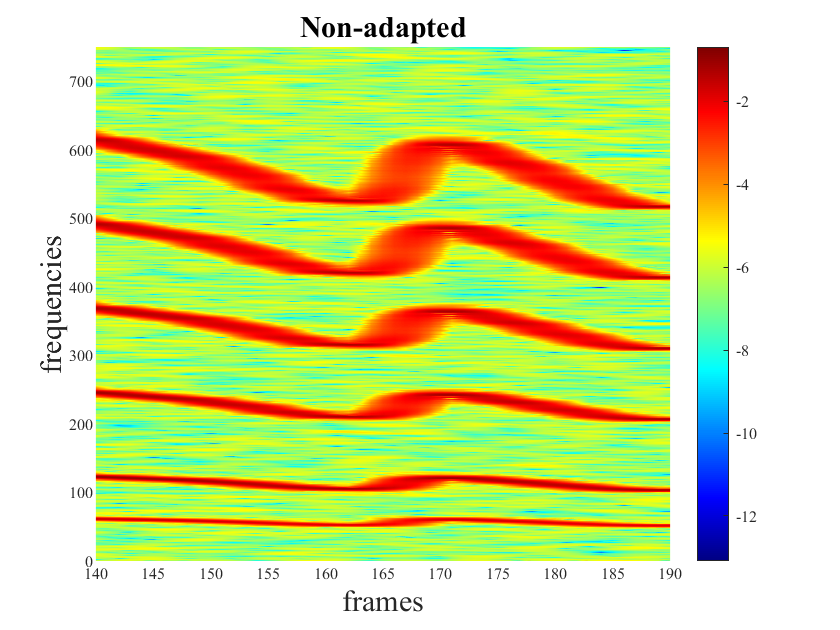}
  \caption{DSTFT.}
\end{subfigure}%
\hspace{-.5cm} \begin{subfigure}{.3\textwidth}
  \centering 
  \includegraphics[width=5cm]{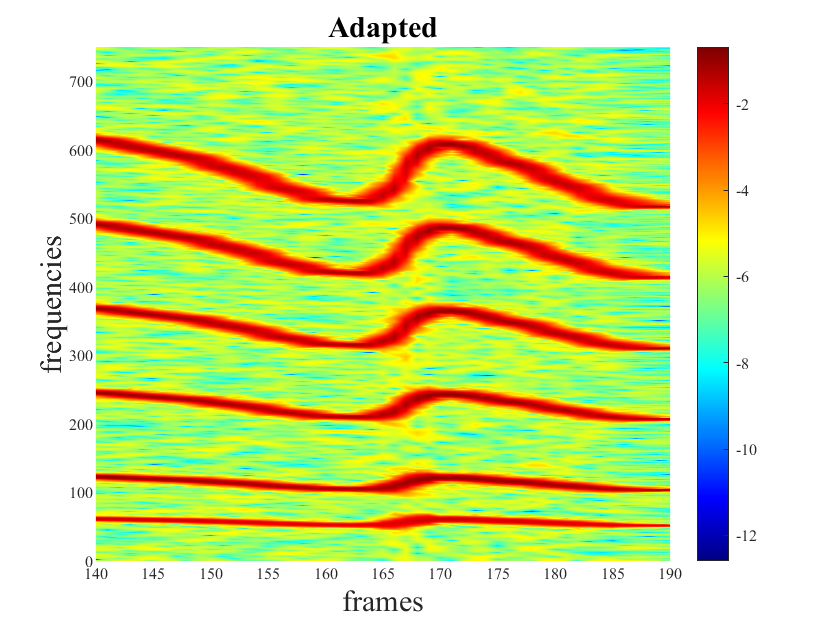}
  \caption{DASTFT}
\end{subfigure}
\caption{Zooms in DSTFT and DASTFT.}
\label{fig::4.2.3}
\end{figure}

The objective is to find a good STFT representation for a multi-harmonic vibration signal where the main harmonic is related to the shaft speed in an aircraft engine. The challenge is then to find a window length able to track fast frequency variations while maintaining a good frequency resolution. For reasons of interpretability, we choose in this experiment only time-varying windows (i.e $\theta_{t,f}=\theta_t$). In Fig. \ref{fig::4.2.1} b) and c) shows the obtained spectrogram and the windows length over time. For comparison purpose, we also show the spectrogram computed from the DSTFT \cite{dstft} minimizing \eqref{eq::eq12} in Fig. \ref{fig::4.2.1} a). In Fig. \ref{fig::4.2.3}, we show zooms in a non-stationary region for both spectrograms. We see that our method adapts to time-varying frequency content of the signal: the window length takes large values in stationary sections and small values in transient sections while constant window length fails to track fast frequency variation.\footnote{It is noteworthy that both spectrograms have the same size according to our definition of STFT.} It is important to note that achieving good frequency and time resolutions is very important in many vibration spectrogram-based applications such as instantaneous frequency estimation \cite{leclere2016multi}.

\section{Conclusion}
\label{sec:5_}
We presented a modification of the DSTFT making this operation adaptive to both transient and stationary components in the same time-frequency representation. We have proposed an adaptation criterion to adapt on-the-fly the window lengths to a given signal. We show through two examples the benefit of using adaptive window lengths with the simplicity of applying gradient descent instead of grid search. 

\clearpage

\bibliographystyle{unsrtnat}
\bibliography{refs}  

\end{document}